\documentclass[twocolumn]{aastex631}

\begin{document}

\title{Comparing the Spatial Correlation of Binary Black Hole Mergers to Large-Scale Structure through the Illustris Simulation}

\author[0009-0001-9017-659X]{Shaniya M. Jarrett}
\affiliation{Department of Physics and Astronomy, Vanderbilt University, Nashville, TN 37240, USA}
\affiliation{Department of Physics, Fisk University, Nashville, TN 37208, USA}
\affiliation{Department of Astronomy, University of Maryland, College Park, MD 20742, USA}

\author[0000-0003-2227-1322]{Kelly Holley-Bockelmann}
\affiliation{Department of Physics and Astronomy, Vanderbilt University, Nashville, TN 37240, USA}

\author[0000-0002-4464-214X]{Robert J. Scherrer}
\affiliation{Department of Physics and Astronomy, Vanderbilt University, Nashville, TN 37240, USA}

\begin{abstract}
Gravitational waves (GWs) have provided a new lens through which to view the universe beyond traditional electromagnetic methods. The upcoming space-based gravitational wave mission, Laser Interferometer Space Antenna (LISA), will give us the first glimpse of the sky in mHz gravitational waves, a waveband that contains a rich variety of sources including massive binary black hole (MBBH) mergers. In this work, we investigate the spatial distribution of MBBH mergers versus the galaxy distribution to determine how well LISA could be used as a unique and independent probe of large-scale structure. We compare the two-point correlation function (2pt CF) of MBBH mergers to that of galaxies within the cosmological hydrodynamic simulation IllustrisTNG.  Our results show that MBBH mergers exhibit stronger clustering than galaxies at scales less than 10 Mpc $h^{-1}$, particularly at higher redshifts, and that the bias is relatively constant as a function of separation. These findings imply that the spatial distribution of MBBH mergers detectable by LISA could inform the observed galaxy distribution. In addition, this implies that searches for a cosmological background in LISA data could use a prior derived from large-scale structure observations to subtract the MBBH foreground.
\end{abstract}

\keywords{gravitational waves—stars:black holes—methods: statistical—galaxies:clusters:general—(cosmology:)large-scale structure of universe—multimessenger astronomy—LISA}

\section{Introduction}

The detection of gravitational waves (GWs) has provided a new lens through which to view the universe, notably through the coalescence of binary black holes~\citep{abbott2016,abbot2018}. These waves are ripples in the curvature of spacetime caused by the extreme accelerative motions of massive objects such as binary neutron stars and black holes (BHs) \citep{einstein1916,abbott2016}. With the Laser Interferometer Space Antenna (LISA), an ESA/NASA space-based gravitational wave mission set to launch in 2035, we will access the mHz gravitational waveband, a domain expected to contain a rich variety of known astronomical sources, including massive binary black hole (MBBH) mergers, white dwarf, neutron star, and stellar-mass black hole binaries, and extreme mass ratio inspirals, as well as the expectation that currently unknown sources may be present and may create a stochastic gravitational wave background \citep{redbook}. LISA is capable of detecting MBBH binaries in the mass range between $10^4 M_\odot $ to $10^8 M_\odot$ out to redshift  \(z=20\) at frequencies from 0.1 mHz to 1 Hz \citep{AWG2023}. 

One of the great benefits of GW astronomy is that it provides complementary information to what is gleaned from electromagnetic observations. Gravitational waves encode distances and masses, which are difficult to obtain directly from electromagnetic observations, at the expense of spatial resolution on the sky.  This makes GWs a potentially valuable tool to address problems at cosmological scales. One such example may be large-scale structure (LSS). While LSS is broadly homogeneous, it features notable inhomogeneities that provide valuable information about gravity, galaxy formation, dark energy, and how these change as the universe evolves \citep{Zehavi2011,karcher2024,hamilton1988,geller1989,hawkins2003,Frenk1991,hubble1926}. Studying the spatial distribution of galaxies serves as a proxy for the distribution of matter on large scales \citep{kaiser1984}, and allows researchers to constrain dark matter properties \citep{ostriker1995}, map galaxy distributions \citep{cole2005}, and select galaxy samples to estimate distances and study the universe's expansion \citep{eisenstein2005}.

Our understanding of LSS has come largely from surveys such as the Sloan Digital Sky Survey (SDSS), and the Dark Energy Survey (DES) where the focus has traditionally been on galaxies \citep{doroshkevich2004,ahumada2020, blanton2017}. However, recent efforts are beginning to explore the clustering of various mergers and their applications to LSS, using GW detectors like LIGO, Cosmic Explorer, and Einstein Telescope. These studies include measuring stellar-mass binary black hole mergers and their relation to LSS \citep{banagiri2020, vijaykumar2023, Peron2024}, analyzing anisotropies in merger distributions \citep{Payne2020}, and using compact object mergers from third-generation detectors to constrain cosmological parameters \citep{Kumar2022,Libanore2021}.

In this work, we investigate the feasibility of probing LSS through MBBH mergers using LISA. As a first step, we compare the distribution of MBBH mergers to that of galaxies using a two-point correlation function (2pt CF) within the cosmological hydrodynamic simulation, IllustrisTNG \citep{pillepich2019}. We also consider bias, which refers to how tracers tend to cluster more strongly compared to the underlying matter distribution. Since mergers are more likely to form in the densest regions of the universe, selecting only MBBH mergers effectively samples from the peaks in these density fields. This naturally leads to a higher correlation function compared to sampling all regions, including those of lower densities. In this case, we would expect that MBBH mergers will appear more clustered than galaxies. Understanding the relationship between these two tracers is foundational to ascertain how well MBBH mergers serve to trace the galaxy distribution, thereby setting the stage to evaluate LISA data as a LSS probe.

The purpose of this work is to address two main areas: the limitations that come from galaxy distribution measurements and their relation to MBBH mergers. First, despite the advanced methods of measuring galaxy locations, EM observations can still introduce bias through scattering within the positioning and clustering statistics, which are crucial for understanding LSS \citep{scelfo2020}. Second, previous work has shown the potential capability for third-generation GW detectors to use stellar-mass black hole mergers as tracers of the underlying galaxy distribution \citep{shao2022}, but the relationship between galaxy distributions and MBBH mergers within the mass range detectable by LISA has received less attention. Also, recent work has provided insights on how stellar mass black hole mergers trace LSS \citep{stiskalek2021,Smith2025}. This paper is a first step in analyzing the utility of gravitational waves as an independent LSS tracer, as well as its potential to provide novel constraints on the $\Lambda$CDM Model \citep{planck2018}. 

This paper is organized as follows: section 2 provides an overview of the simulation, the 2pt CF, and our methodology. Section 3 presents our results for the 2pt CF of galaxies versus MBBH mergers up to cosmic noon. Section 4 contains a discussion of these results and next steps, and section 5 summarizes our findings.

\section{Methods}

\subsection{Illustris Simulation and Galaxies}
 
Our data comes from the publicly-available cosmological hydrodynamic simulation suite {\it Illustris} ~\citep{nelson2015,vogelsberger2014}. These simulations evolve dark matter, stars, gas, and massive black holes, and incorporate key astrophysical processes with subgrid models, including gas cooling and photo-ionization, star formation, stellar evolution, and BH feedback. Gas cooling is modeled based on density, metallicity, and temperature, while star formation follows the Kennicutt-Schmidt law. Stellar evolution accounts for mass return from various types of stars, and BH feedback is included with mechanisms for both quasar-mode and radio-mode activity, which influences the surrounding gas and modifies the cooling rates. 

We use data from the Illustris-3 simulation of a volume with box length of \( 75~\text{Mpc}~\textit{h}^{-1} \). The average gas cell mass is $0.0057 \times 10^{10} \, M_{\odot}/h$ and dark matter particle mass is $0.0282 \times 10^{10} \, M_{\odot}/h$. In this work, we focus on mergers between $z=0-2$. We limited the redshift to $z=2$ because this marks the peak of galaxy formation, and to ensure a large enough number of mergers when comparing their 2pt CFs. Table \ref{tab:simulation_snapshots} includes details used to constrain the data in Figure \ref{fig1}. This lists redshift ranges of  \(z=0 - 0.20\), \(z=0.20 - 0.46\), \(z=0.46 - 0.85\), and \(z=0.85 - 2.0\), the number of galaxies ranging from about $4 \times 10^{5}$ to $6 \times 10^{5}$, and the number of mergers for each range which are over 1600. The following cosmological parameters \citep{planck2016} are used in the simulation: $\Omega_m = 0.3089$, $\Omega_{\Lambda} = 0.6911$, $\Omega_b = 0.0486$, and $H_0 = 100h \, \text{km s}^{-1} \text{Mpc}^{-1}$ with $h = 0.6774$.

In Illustris, a subhalo can host a single galaxy, be part of a larger galaxy group, or contain no visible galaxy at all if it has been stripped of its baryonic content or failed to form stars efficiently. The presence of a galaxy within a subhalo is determined by the concentrations of stars and gas, and therefore stellar mass. 
Dark matter halos are identified with the SUBFIND algorithm \citep{springel2001}. We select galaxies based on stellar mass between $10^9 M_\odot - 10^{12.5} \, M_\odot$
\citep{rodriguez-gomez2016,haslbauer2019,pillepich2019}.

\begin{table*}[ht]
\centering
\caption{Selected Simulation Data}
\label{tab:simulation_snapshots}
\small  
\begin{tabular*}{0.7\textwidth}{@{\extracolsep{\fill}}ccc}
\hline
\textbf{Redshift}  & \textbf{MBBH Mergers [\#]} & \textbf{Galaxies [\#]} \\
\hline
  0.00 - 0.20   & 1631 & 425245    \\
  0.20 - 0.46   & 1654 & 441553    \\
  0.46 - 0.85   & 1585 & 453537    \\
  0.85 - 2.00   & 1633 & 636396     \\
\hline
\end{tabular*}

\end{table*}

%

The specific algorithm used to identify halos and subhaloes containing galaxies includes the Friends-of-Friends (FoF) algorithm. This algorithm identifies gravitationally bound dark matter particles, linking particle groups by a specific distance of 0.2 times the mean interparticle separation. This process recursively groups particles to form clusters until a dark matter halo is defined that is gravitationally bound. Within these halos, the subhalo fields are derived with the Subfind algorithm, which detects subhalos by reviewing the gravitational binding of gas and stars \citep{illustris2018,degraf2020,sijacki2015}. By identifying areas of over-density within the FoF groups, then assigning nearby baryonic particles to these regions, it can determine if there are potential galaxies within the subhalos. Subfind also looks at the thermal energy of the gas to help to confirm if baryonic particles are actually bound to the subhalo.

\subsection{Illustris Massive Binary Black Hole Mergers}

A massive BH is initially seeded with a mass of $10^{5} h^{-1} M_{\odot}$ at the center of a dark matter halo once its mass exceeds $5 \times 10^{10} h^{-1} M_{\odot}$, as long as it does not already contain a BH particle.  This approach is intended to correspond to direct collapse black hole seed formation models, where massive BHs form rapidly in the densest regions of protogalaxies due to the gravitational collapse of primordial gas \citep{haehnelt1993,bromm2003,begelman2006}. The first black holes here form between \(z=10 - 11\).

Once seeded, BHs can increase in mass through Bondi-Hoyle gas accretion and by merging with other BHs, so the BH phase space and the properties of the immediate gas reservoir are tracked at each step in the simulation, as well as the mass accreted and the time of any merger. A merger is defined when two BHs come within a softening length of 1 kpc. This does not depend on relative velocity, due to the poorly constrained black hole velocities from scattering interactions in the halos \citep{sijacki2015}.


This separation is still about a billion times larger than the true merger and ignores the complex physics of few-body scattering, inspiral through gaseous disks, and gravitational radiation that would prolong the true merger for millions to billions of years \citep{khan2021,khan2024}.   

We focus on MBBH mergers between \(2.84 \times 10^{5} \, M_{\odot}\) to \( 10^{8} \, M_{\odot}\) as this is the smallest merger mass in Illustris for our redshift ranges, and it falls within the detectable mass range for LISA, with a total number of about 6500 mergers. This total is split such that each redshift range and the number of mergers per bin are roughly equal.

\vspace{4mm}
\subsection{Correlation Function}


To quantify LSS, we use the 2pt CF ${\xi}(r)$, which gives the excess probability of finding a pair of objects at a given separation compared to a random distribution:
\begin{equation}
{\xi}(r) = {\frac{DD(r)}{RR(r)} - 1},
\label{eq:Basic_2pt_cf}
\end{equation}
\noindent where \( DD(r) \) is the number of pairs at separation \( r \) from our data, and \( RR(r) \) is the number of pairs if the objects were randomly distributed. It is well known that the 2pt CF can constrain galaxy formation and the parameters in $\Lambda$CDM, and ${\xi}(r)$ has been calculated for various astrophysical objects within the Illustris simulation \citep{fontanot2024,sanchez2017,springel2018,Zhang2023UsingTT}, but not for MBBH mergers.
There are several ways to calculate the 2pt CF \citep{landy1993,davis1983,zhao2023}, many of which account for edge effects of a finite survey volume \citep{yue2024,yuan2023, breton2021}. The method we use is based on the Landy-Szalay prescription:
\begin{equation}
{\xi}(r) = \frac{\text{DD}(r) - 2\text{DR}(r) + \text{RR}(r)}{\text{RR}(r)},
\label{eq:Improved_2pt_cf}
\end{equation}
\noindent where we build from Equation~\ref{eq:Basic_2pt_cf} and now have \( DR(r))\), which is the number of pairs between data and random distributions, and is calculated using \texttt{Corrfunc} \citep{sinha2019}. Corrfunc is optimized for speed by organizing data points and vectorizing them to ensure cache locality to facilitate rapid memory access and parallelization. It has been used for tasks such as measuring 3D clustering of SMBH-hosting halos, projected galaxy clustering for halo models, and angular clustering of galaxies \citep{singh2023,yuan2025,to2021}.

\begin{figure*}[ht!]
    \centerline{\includegraphics[width=.85\textwidth, keepaspectratio]{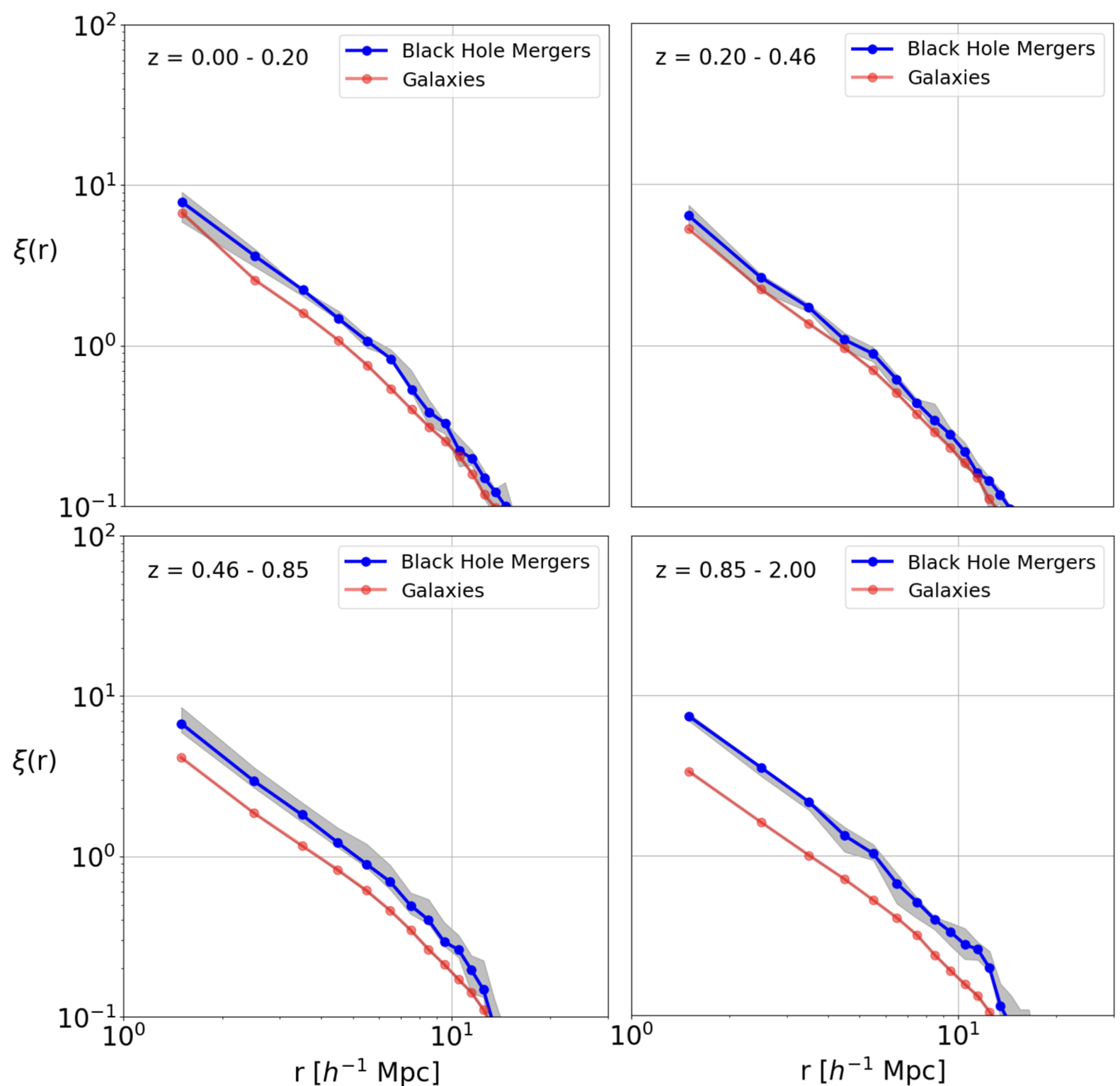}}
    \caption{The 2pt CF of MBBH mergers (blue) and galaxies (red) in Illustris-3 at redshifts  \(z=0 - 0.20\), \(z=0.20 - 0.46\), \(z=0.46 - 0.85\), \(z=0.85 - 2.0\). We see that both MBBH mergers and galaxies are more clustered at smaller scales ($< 5\,\mathrm{Mpc}\  h^{-1}$
), and that MBBH mergers consistently exhibit stronger clustering compared to galaxies for all scales, particularly for higher redshift ranges.}
   \label{fig1}
\end{figure*}

We first constrain the merger and galaxy masses as mentioned in sections 2.1 and 2.2, before the 2pt CF calculation. We choose 20 logrithmically-spaced bins in separation and selected redshift ranges to include a relatively equal number of mergers, with around 1600. The bins range from \(1\) to \( 30~\text{Mpc}~\textit{h}^{-1} \), and include an rmax up to 20 Mpc $\textit{h}^{-1}$ to minimize edge effects in our volume. These values are based on the Corrfunc package recommendations, and are maintained to facilitate a direct comparison between MBBH mergers and galaxies \citep{sinha2019software}. We calculate the 2pt CF in the following redshift bins: \(z=0 - 0.20\), \(z=0.20 - 0.46\), \(z=0.46 - 0.85\), \(z=0.85 - 2.0\), and derive confidence intervals using bootstrapping methods.

\section{Results}

\begin{figure*}[ht!]
    \centerline{\includegraphics[width=.9\textwidth, keepaspectratio]{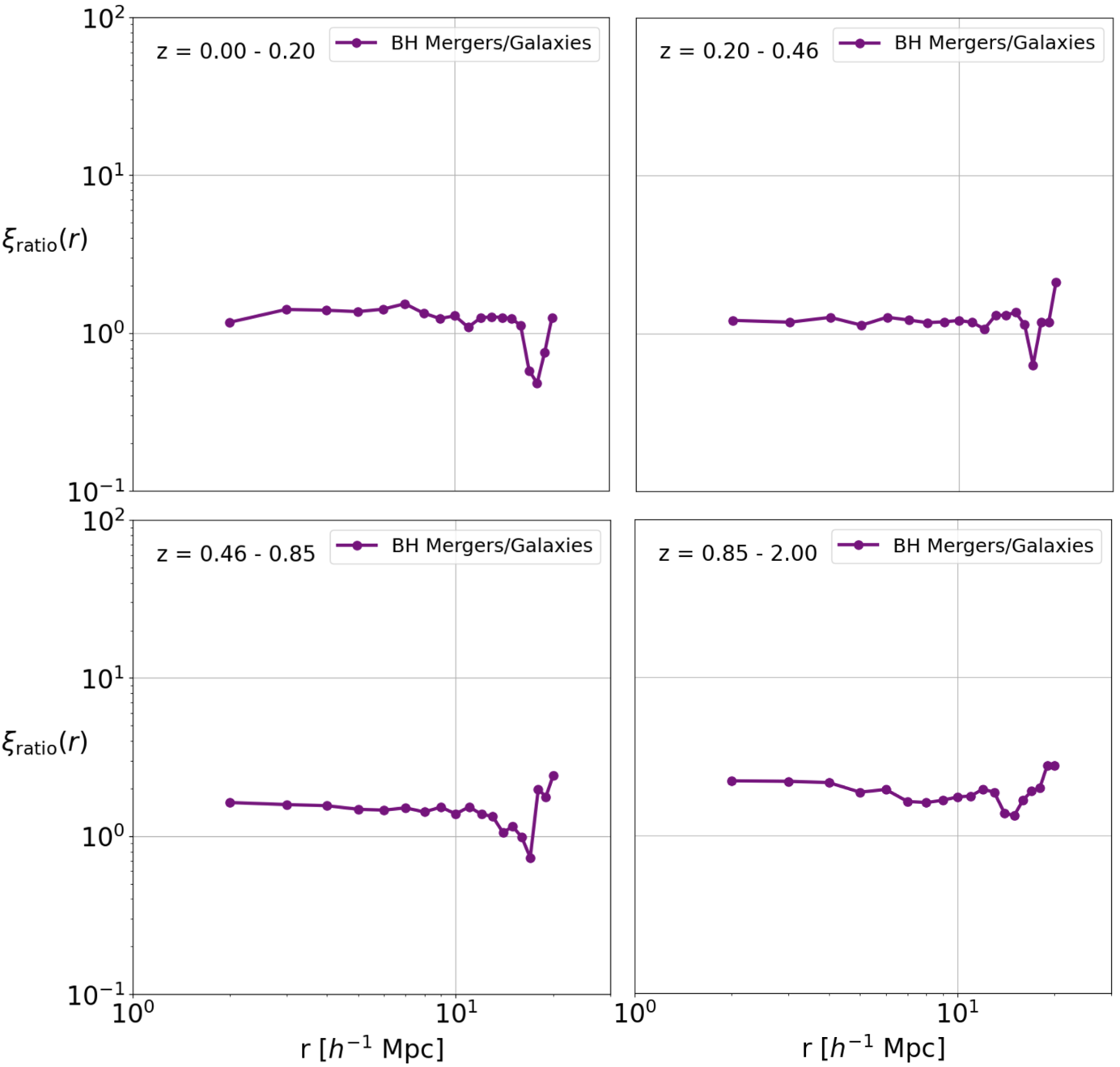}}
    \caption{The 2pt CF bias for MBBH mergers versus galaxies (purple) at redshifts  \(z=0 - 0.20\), \(z=0.20 - 0.46\), \(z=0.46 - 0.85\), \(z=0.85 - 2.0\). For all redshift ranges, the bias remains relatively consistent at scales below  $ 15\, \mathrm{Mpc} \ h^{-1}$ with more variation at larger distances. There is a slight increase in the ratio observed at higher redshifts.}
   \label{fig2}
\end{figure*}

Figure \ref{fig1} shows the 2pt CF for MBBH mergers and galaxies in four redshift bins. In the lowest redshift bin \(z = 0.0 - 0.20\), we see that MBBH mergers exhibit stronger clustering around 1 Mpc \(h^{-1}\) compared to galaxies. Both correlation functions follow a power-law slope, $\xi(r) = \left( \frac{r_0}{r} \right)^{\gamma}$, with $\gamma_{\mathrm{m}} = 2.28$
 for the mergers and $\gamma_{\mathrm{g}} = 2.07$ for the galaxies. The stronger clustering at smaller scales supports findings from other studies, which suggest that MBBH mergers preferentially form in dense galactic environments. With redshift ranges \(z = 0.20 - 0.46\), the clustering patterns of MBBH mergers and galaxies become more similar compared to smaller redshift ranges at slopes of $\gamma_{\mathrm{m}} = 2.0$ and $\gamma_{\mathrm{g}} = 2.05$. Both objects exhibit a convergence in their spatial distribution, and the similarity in clustering strength implies that the environments where MBBH mergers and galaxies form and evolve are more similar.

For redshift range \(z = 0.46 - 0.85\), MBBH mergers continue to show higher clustering strength than galaxies at smaller scales and have slopes of $\gamma_{\mathrm{m}} = 2.1$ and $\gamma_{\mathrm{g}} = 2.04$. However, both types of objects exhibit a decrease in clustering strength as we move to larger scales. The decreasing clustering strength at larger radii aligns with the general understanding of the cosmic structure evolving towards homogeneity at large scales. Lastly, at redshift ranges \(z = 0.85 - 2.0\), the clustering patterns of MBBH mergers and galaxies become more distinct at $\gamma_{\mathrm{m}} = 2.06$ and $\gamma_{\mathrm{g}} = 2.03$. MBBH mergers show consistently higher clustering behaviors later demonstrated in Figure \ref{fig2}. This pattern supports the hypothesis that the formation of MBBH mergers is closely linked to regions with high galaxy densities. 

Our results show that MBBH mergers exhibit stronger clustering than galaxies across all redshift bins, indicating a higher likelihood of occurring in over-dense environments compared to galaxies. This indirectly suggests that these mergers preferentially form in galaxy-dense regions, which may play a role in driving MBBH evolution. Additionally, within these environments, MBBH mergers are more likely to be found in close proximity to one another compared to the general galaxy population. This proximity increases the frequency of galaxy interactions and can lead to a higher chance of merger events.

The calculated confidence intervals for the Corrfunc 2pt CF are represented by the shaded regions in Figure \ref{fig1}, which indicate a high level of precision in our measurements. The tight clustering of data points around the fitted line within the 95\%\ confidence interval suggests reliable correlation functions for both MBBH mergers and galaxies, and have an average uncertainty of \(\Delta \xi / \xi  \approx 12.8\%\) for the mergers, and \(\Delta \xi / \xi \approx 0.7\%\) for the galaxies. This result was calculated by taking the half-width of the 95\%  confidence interval at each separation and dividing it by the corresponding \(\xi(r)\) value, then averaging across all scales where \(\xi(r) > 10^{-1}\).  Our results are consistent with the general trend that clustering strength decreases as radius increases, and that the slope of the clustering for both MBBH mergers and galaxies, which ranges from $\gamma_{\mathrm{m}} = 2.03 - 2.07$, and $\gamma_{\mathrm{m}} = 2.0 - 2.28 $, increases with time. Even when accounting for uncertainties, the correlation functions for MBBH mergers and galaxies remain statistically distinct across redshifts, with minimal overlap.

In Figure \ref{fig2}, we show the ratio of the 2pt CF for MBBH mergers relative to galaxies across different redshift ranges. This ratio provides insight into the clustering bias of MBBH mergers compared to the underlying galaxy distribution, illustrating how their spatial distributions evolve over time. Starting with the top row, for the redshift range \(z = 0.0 - 0.20\), the ratio remains slightly above 1 for  \( 1\,\text{ Mpc}\ h^{-1}  < r < 10\,\text{ Mpc}\ h^{-1} \). At larger scales beyond 10 Mpc \(h^{-1}\), the ratio begins to fluctuate due to statistical noise driven by low pair counts at large separations. This suggests that these features are artifacts of edge effects or limited statistics rather than the clustering behavior.

For \(z = 0.20 - 0.46\), the bias ratio remains closer to 1, again with some variance at larger scales. For the bottom row, the redshift range \(z = 0.46 - 0.85\) has a ratio that is slightly higher than in the lower redshift bins, approaching a value just under 2, with similar fluctuations for r $>$ 10 Mpc \(h^{-1}\). At the highest redshift range of \(z = 0.85 - 2.0\), the bias ratio is the highest compared to the other ranges, exceeding a value of 2. The bias remains relatively flat across all redshifts, suggesting that the slope of the MBBH merger 2pt CF is comparable to that of galaxies. This implies that MBBH mergers can be used to trace the shape of the galaxy correlation function, even if their clustering amplitudes differ. However, their clustering remains comparable to that of galaxies, suggesting that MBBH mergers trace LSS with minimal clustering bias and may serve as suitable tracers through LISA.

\section{Discussion}

This work contributes in providing a novel method for exploring LSS by comparing the spatial clustering of MBBH mergers and galaxies. By establishing that MBBH mergers have a consistent clustering bias relative to galaxies through the 2pt CF, we gain further insight into the clustering behavior of the these mergers, and demonstrate that the mergers trace the underlying shape of the universe. This implies that future gravitational wave detectors such as LISA, could serve as independent tracers of LSS, alongside traditional galaxy surveys.

Although our results show that mergers are more clustered across all scales, both mergers and galaxies show stronger clustering for smaller distances, which is consistent with clustering patterns at lower redshifts \citep{yang2019}, general galaxy clustering behavior \citep{Wang2013,basilakos2004}, and 2pt CFs performed on similar galaxy mass ranges \citep{Bose2023,fontanot2024}. The similarity in clustering between MBBH mergers and galaxies is most prominent at lower redshifts, specifically within the range of
 \(z = 0 - 0.46\). The slopes for the galaxies based on redshift, ranged from $\gamma_{\mathrm{m}} = 2.03 - 2.07$, while the slopes of MBBH mergers ranged from $\gamma_{\mathrm{m}} = 2.0 - 2.28 $. These values are a bit steeper compared to the dark matter halo correlation slope of  $\gamma_{\mathrm{m}} = 1.8$ \citep{yang2005}, suggesting higher clustering for both populations compared to the underlying dark matter distribution. Additionally, lower clustering behavior prevalent at higher redshifts suggests a lower number of mergers and could indicate a decrease in merger rates. \citep{vanSon2022}. We see this in the Illustris simulation, as there are a total of 1255 mergers for redshifts \(z=1 - 2\), compared to 4417 mergers for redshifts \(z= 0 - 1\).

Since the 2pt CF quantifies clustering at different scales and allows us to trace the underlying matter distribution, avoiding scale-dependant bias ensures that those measurements reflect the true structure more reliably. In finding a linear merger–galaxy bias, we can simplify both the interpretation of merger clustering and the subtraction of the MBBH foreground in preparation for LISA. In the instance where there is scale-dependent bias between galaxies and matter, determining cosmological parameters becomes more challenging. Similarly, if the bias between mergers and galaxies were scale-dependent, removing the foreground would require fitting a separate bias model at each scale. Additional scale-varying terms would need to be introduced into the two-point correlation function to recover unbiased constraints \citep{Desjacques2018}. However, we have shown a linear bias, which implies that MBBH mergers trace LSS in the same way as galaxies over all separations. A scale-independent bias allows us to infer the amplitude of matter fluctuations directly, without resorting to corrections that inflate parameter uncertainties \citep{PhysRevD.95.023505, PhysRevD.100.043514}. This improvement in LSS measurement accuracy could also help constrain $\Lambda$CDM values, support CMB studies that rely on cross-correlation with LSS tracers, and expand on the formation and evolution of galaxies and objects within them by connecting MBBH merger events to their surrounding halo properties and galaxy environments.

A few caveats to note pertain to the Illustris simulation. In using a box size of \( 75~\text{Mpc}~\textit{h}^{-1} \), we limit our ability to fully capture LSS, and increases the uncertainty from cosmic variance. This could have an impact on the 2pt CF as we cannot probe these larger scales. Another challenge to consider is the merging time of the MBBHs, as the Illustris simulation simplifies the post-dynamical friction evolution of mergers and omits the prolonged MBBH inspiral once they reach a softening length of 1 kpc. Earlier work has shown that merger rate predictions relevant to LISA, are sensitive to assumptions about gas accretion efficiency, dynamical friction models, and host galaxy dynamics \citep{holleybockelmann2010,chen2022,ni2022astrid}. Selection bias should also be noted as the simulation may over or underestimate merger rates in different environments and make assumptions about how clustering evolves with redshift. This could potentially skew the observed clustering behavior. Outside of the simulation, we note that since we expect MBBH mergers to preferentially form in the densest regions of the universe, selecting only mergers might introduce a form of peak bias. Sampling from these high-density environments could lead to a higher clustering statistic compared to galaxies, which span a broader range of environments.

That being said, our results show promise towards inferring LSS from MBBH merger measurements. They also offer insight into the formation and evolution of mergers, galaxies, and can provide insight into the relationship between MBBH mergers and their host galaxies \citep{IzquierdoVillalba2023}, alongside the expansion history of the universe. Knowing this information, we can use MBBH mergers as independent tracers of LSS and avoid EM bias, giving more accurate LSS measurements of the underlying matter distribution. 

Additional steps could be taken to improve on the accuracy of our results, determine how well LISA will be able to resolve these sources, and establish the number of mergers needed to infer LSS. Regarding accuracy, incorporating dynamical friction and inspiral time as seen by LISA could improve precision \citep{banks2022,Li2022,langen2024}. To assess resolving sources, a next step could involve creating a simulated map of MBBH merger locations based on LISA’s sensitivity, where each source is represented by a probabilistic sky localization region rather than a fixed point. The 2pt CF could then be recomputed from these probabilistic regions using a Monte Carlo method to evaluate how well the correlation results are consistent with the MBBH correlations found here. If there is concurrence, it quantifies how we can use LISA observations to constrain LSS. From there, further work would be needed to determine the minimum number of LISA merger observations required to reliably reconstruct LSS.

\section{Conclusion}

In this work, we explored the spatial correlation of MBBH mergers through the Illustris simulation, focusing on their clustering behavior compared to galaxies. Using the 2pt CF, we analyzed the clustering patterns of these mergers and galaxies up to cosmic noon. Our results indicate that MBBH mergers exhibit stronger clustering at smaller scales compared to galaxies, across all redshift ranges of \(z=0\) to \(z=2\), and that the bias remains relatively flat across all redshifts.

The redshift ranges where the MBBH merger clustering was the most similar to galaxies was for \(z=0.20\) to \(z=0.46\) with slopes of \(\gamma_{\mathrm{m}} = 2.0\) and \(\gamma_{\mathrm{g}} = 2.05\). The most significant difference was shown at ranges \(z=0.85\) to \(z=2.0\), where the clustering patterns of MBBH mergers and galaxies become more distinct, with MBBH mergers consistently showing higher clustering behavior at slopes of \(\gamma_{\mathrm{m}} = 2.28\) compared to \(\gamma_{\mathrm{g}} = 2.03\). For all bins, our measured slopes range from \(\gamma_{\mathrm{m}} = 2.0 - 2.28\) for mergers and \(\gamma_{\mathrm{g}} = 2.03 - 2.07\) for galaxies, which are slightly steeper compared to the dark matter halo correlation slope of \(\gamma \sim 1.8\). 
 
The bias ratios further support these findings, showing that the MBBH mergers demonstrated slightly stronger clustering than galaxies, with the largest deviation occurring at \(z = 0.85 - 2.0\), and the smallest being at \(z = 0.20 - 0.46\). The confidence intervals calculated for the 2pt CF that are represented by the shaded regions in Figure~\ref{fig1}, show the 95\%\ confidence intervals and indicate a high level of precision in our measurements. Over the range where \(\xi(r) > 10^{-1}\), the average uncertainty is \(\Delta \xi/\xi \approx 12.8\%\) for MBBH mergers and 0.7\% for galaxies.

The slope of the 2pt CF demonstrates how clustering strength decreases with distance. This observed trend of decreasing clustering strength with increasing radius is consistent with the general understanding of cosmic structure evolving towards homogeneity at large scales. However, we find that mergers are more clustered at all ranges compared to galaxies. This suggests that MBBH mergers are more likely to occur in dense, galaxy-rich environments, which may play a role in driving their evolution. These environments may also increase the likelihood of galaxy interactions and BH merger events. While the amplitude of the bias illustrates stronger clustering of MBBH mergers, the relatively linear behavior with radius across all redshifts suggests that MBBH mergers and galaxies trace the same underlying structure.

Our findings suggest that galaxies can serve as a proxy for MBBH mergers in mapping LSS, providing a method for studying the universe's structure through GWs. This has notable implications for future space-based gravitational wave missions like the LISA, which could use the spatial distribution of MBBH mergers to gain insights into the formation and evolution of galaxies and broader large-scale structures. By advancing our understanding of the spatial distribution of MBBH mergers and their relationship to galaxy distributions, this work helps set the stage for future gravitational wave missions as a probe of LSS.

\vspace{5mm}

\subsection{Acknowledgments}
Funding for this work has been provided by the National Science Foundation (NSF 2125764). We thank Bill Smith at Vanderbilt University for his insights and Manodeep Sinha for guidance setting up Corrfunc.


\bibliography{Citations}
\bibliographystyle{aasjournal}

\end{document}